# Understanding the Strength of the Selenium - Graphene Interfaces

Vidushi Sharma[1], David Mitlin[2,*], Dibakar Datta[1,*],

[1]Department of Mechanical and Industrial Engineering, New Jersey Institute of Technology, Newark, NJ 07103, USA

[2] Materials Science and Engineering Program & Texas Materials Institute, The University of Texas at Austin, Austin, TX, 78712-1591, USA

**KEYWORDS:** Density Functional Theory (DFT), Selenium, Graphene, Silicon, Interface, Charge Transfer, Density of States (DOS)

## Abstract

We present comprehensive first-principles Density Functional Theory (DFT) analyses of the interfacial strength and bonding mechanisms between crystalline and amorphous selenium (Se) with graphene (Gr), a promising duo for energy storage applications. Comparative interface analyses are presented on *amorphous* silicon (Si) with graphene and crystalline Se with conventional aluminum (Al) substrate. The interface strength of *monoclinic* Se (0.43 J/m$^2$) and *amorphous* Si with graphene (0.41 J/m$^2$) is similar in magnitude. While both materials (*c*-Se, *a*-Si) are bonded loosely by van der Waals (vdW) forces over graphene, interfacial electron exchange is higher for *a*-Si/graphene. This is further elaborated by comparing potential energy step and charge transfer (Δq) across the graphene interfaces. The Δq for *c*-Se/Gr and *a*-Si/Gr are 0.3119 $e^{-1}$ and 0.4266 $e^{-1}$, respectively. However, interface strength of *c*-Se on 3D Al substrate is higher (0.99 J/m$^2$), suggesting stronger adhesion. The *amorphous* Se with graphene has comparable interface strength (0.34 J/m$^2$), but electron exchange in this system is slightly distinct from *monoclinic* Se. The electronic characteristics (density of states analysis) and bonding mechanisms are different for *monoclinic* and *amorphous* Se with graphene and they activate graphene via surface charge doping divergently. Our findings highlight the complex electrochemical phenomena in Se interfaced with graphene, which may profoundly differ from their "free" counterparts.

# 1. Introduction

The development of 'next-generation' electrodes by combining materials into composite structures is gaining attention to enhance the energy and power-density of existing battery technologies. Two or more materials are amalgamated in varied nano- and micro-structures, where each component can contribute in one or many ways as an active electrode[1, 2], composite additive and a binder[3-10], porous matrix[11], or even a current collector[12, 13]. However, much less focus has been directed towards the interface chemistry of these materials. To this end, silicon (Si) is an exemplar anode where issues of cycle life, capacity, volume expansion, and surface reactivity have been successfully addressed by nano-engineering strategies such as Si alloys[14], Si film composites[15], Si and carbon (C) nanoparticles composites[16], and porous Si mixed with carbon-based nanostructures[17]. Naturally, the interface of materials in such systems becomes the focal point, which dictates their applicatory success.

Comparable to alloying Si anode, which has the very high specific capacity ( 3000 mAhg$^{-1}$ ), elemental sulfur (S) cathode can also deliver the high specific capacity of 1675 mAhg$^{-1}$ with its projected energy density being two to three times higher than conventional cathodes[18]. However, the primary concern is the solubility of Li-S reaction intermediate long-chain polysulfides that diffuse into electrolyte, causing shuttle effects[19]. As these problems with Li-S batteries become increasingly prominent[20-22], S is being replaced by heavier chalcogens such as selenium (Se) and tellurium(Te)[23-26]. Currently, Te usage is limited by its higher cost, elemental toxicity, and extremely fast capacity decay due to pulverization[25]. Thus, Se is replacing S as a favored choice. Apart from being less reactive and free from shuttle effects[27], it possesses superior electrical conductivity (1X10$^{-3}$ S.m$^{-1}$ for Se and 5X10$^{-28}$ S.m$^{-1}$ for S)[28] and lithiation rates than Li-S cathode[29]. Furthermore, Se is capable of moving battery technology a step further towards sustainable sodium ion batteries with better sodiation kinetics[30]. This makes Se an ideal candidate for next-generation energy storage materials.

Chalcogens directly react with Li/Na and undergo a conversion-type reaction accompanied by significant volume expansions causing chemo-mechanical degradation. To overcome this, micro and mesoporous C has been used as an additive, first to S[31, 32] and now to Se[33]. This porous matrix of C provides a buffer space for the active electrode Se to expand at ease and continuity of electronic contact. In return, electronegative Se with its large pool of $d$ electrons and polarizability manipulates the surface chemistry of the embedded C matrix and activates it to provide additional Li storage sites[27, 33-35]. As the experimental techniques to infuse Se into C and resultant micro-structures are varied, electrochemical outcome of Se-C cathode is impacted significantly. Electrochemical activity and cycle life of Se-C improves when morphology of C is shifted towards more refined nanostructures such as nanofibers[35] and graphene[36-38]. Therefore, in the latest studies, porous C is now being replaced by graphene in Se-C systems.

Being a 2D derivative of graphite, the most commercialized anode of LIB, graphene independently retains competence to store Li/Na[1, 2]. This is evident in an experimental work by Han et al.[39] in which Se nanoparticles, embedded in a mixture of mesoporous C and graphene, exhibited better discharge capacities and cycle life as LIB cathodes than Se in porous C alone. In a recent experimental and computational study on Si over graphene substrate by Basu et al.[12], slipperiness of graphene surface proved to be effective in combating stresses in Si anode upon lithiation, thereby increasing the cycle life of the electrode. Interface adhesion between Si and the substrates was the primary determinant of electrode cycle life. While many prior studies claim that high adhesion between active electrode material and additive will be beneficial for battery cyclability[40-42], this study proves that low interface adhesion due to slippery graphene surface could be instead more favorable for the battery life. A latest report [43] suggest high interface strength between two materials can cause formation of structurally-disconnected aggregates within electrode. This condition could be avoided if the interface strength between two materials is carefully adjusted along with other physio-chemical factors.

To this end, the present study theoretically investigates the interface between 3D/2D Se/graphene and their future potential in battery applications as composite electrodes. Se-graphene-based works are still in infancy, with most of them being experimental reports. The atomic-level detailed investigation of the Se-graphene interface in terms of interfacial strength, bonding, and overall electronic character is missing. We also present a comparative investigation of amorphous Si/graphene interface as its efficacy is well utilized in the batteries[3-7] and can act as a baseline in this work. The novelty of the present study is: We have determined differences in the interface strength of monoclinic and amorphous Se with the 2D hexagonal lattice of graphene. Se comes in several allotropic forms: monoclinic, trigonal, and amorphous. Being temperature and pressure-sensitive, it undergoes phase transformations during its applications, which remain less understood due to the marginal difference between structures of its different allotropes[44]. Nevertheless, even these marginal structural changes in Se cause fluctuations in interface strength, bonding, the directionality of electron flow, and potential gradient at Se-based interfaces. Furthermore, these characteristics also influence the electronic states of Se and graphene distinctively, which we have investigated in detail using the density of states (DOS) analysis. Lastly, we discuss the consequences of our interface analysis on the application of Se-graphene systems in batteries.

## 2. Computational details

Crystalline(*c*-) and amorphous(*a*-) phase of Se were modeled (fig. 1) before the interface analysis. Monoclinic Se with eight-membered monomer rings $S_8$ was opted as *c*-Se. The latter has structural parameters such as interatomic bond lengths, bond angles, and dihedral angles similar to its other crystalline allotropes[44]. Starting from *c*-Se, amorphous selenium (*a*-Se) was derived by computational quenching [45, 46]. The quenching process required ab initio molecular dynamics (AIMD) within the DFT framework in Vienna Ab initio Simulation Package (VASP)[47]. We performed systematic heating, cooling, and equilibration of Se for 5000 MD time steps with 1 fs time interval under the NVT canonical ensemble. The highest temperature considered (5000 K) was far above the melting point of Se. The final amorphous structure was obtained via DFT optimization of the room temperature AIMD simulated lowest energy (local minima) structure. Projector-augmented-wave (PAW) potentials were used to mimic inert core electrons, while the valance electrons were represented by plane-wave basis set.[48, 49] Plane wave energy cut-off and convergence tolerance for all relaxations were 550 eV and $1.0\times10^{-6}$ eV, respectively. The GGA with the PBE exchange-correlation function was taken into account[50]. For optimizing the initial structures of bulk *c*-$Se_{64}$ and *a*-$Se_{64}$, gamma centered 4 X 4 X 4 k-meshes were employed for good convergence. The energy minimization was done by conjugate gradient method with Hellmann-Feynman forces less than 0.02 eV/Å. The final *c*-Se and *a*-Se phases were identified using radial distribution function (RDF) plots. The RDF plot for *c*-Se in fig.1-a2 presents more than one prominent peak, which is symbolic of crystallinity, and the nearest Se-Se distance of ~2.4Å is noted. In contrast, RDF plot in fig.1-b2 for *a*-Se exhibits nearest Se-Se distance of ~2.38Å and only one prominent peak with low intensity. These Se structures were then interfaced with a graphene (Gr) for further analysis (depicted in fig.1-a3 and -b3).

Vacuum interface model[51] with added vacuum in z-dimensions (normal to graphene plane) was used to calculate the interface energies (illustrated in fig. 2). In total, we studied four interface systems: (*a*- and *c*-)$Se_{64}$/Gr, *a*-$Si_{64}$/Gr, and a 3D/3D *c*-$Se_{64}$/Al for comparison purposes. Here 64 is the number of Se or Si atoms in bulk. For Se-Gr interface systems, Gr comprised of 112 $sp^2$ carbon atoms arranged in a hexagonal lattice, which takes into account the surface area of *c*-$Se_{64}$ crystal in (001) direction. Preparation of *a*-Si/Gr interface system has been reported previously[12] with the number of carbon atoms reduced to 60. In *c*-$Se_{64}$/Al system, monoclinic $Se_{64}$ was interfaced with four atomic layers of Al as substrate. These structures are periodic in x-y dimensions. For DFT calculations, gamma centered 4 X 4 X 1 k-meshes were employed and GGA functional was inclusive of vdW correction to incorporate the effect of weak long-range van der Waals (vdW) forces[52]. All calculations were done with optPBE functional within vdW-DF-family[53, 54].

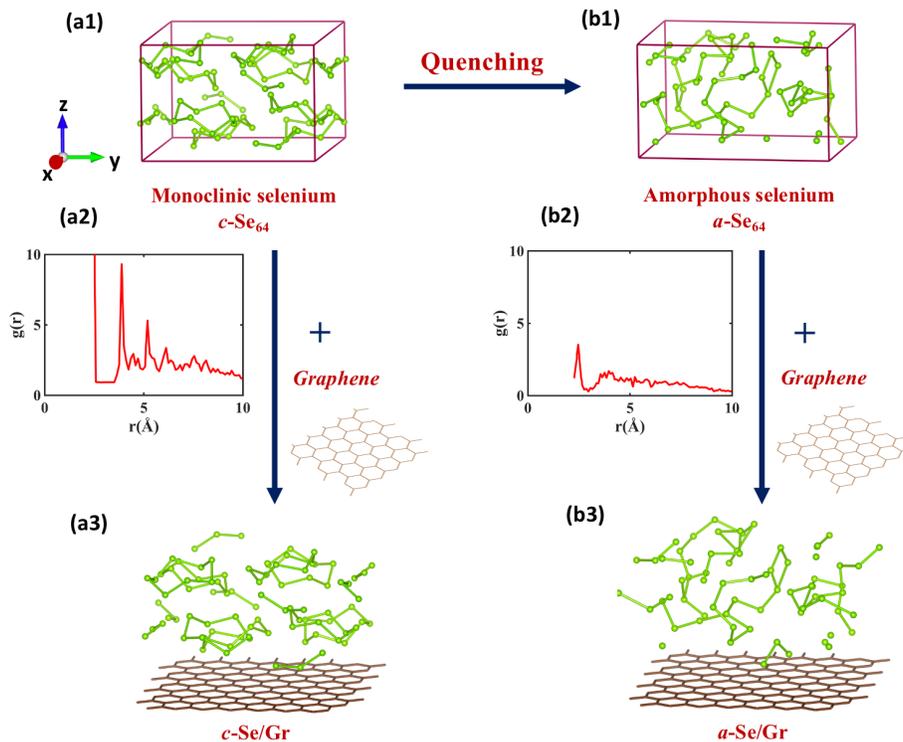

**Figure 1 Schematic illustration of quenching process to generate initial Se structures. (a1)** Optimized monoclinic Se (*c*-Se) having $Se_8$ rings. **(b1)** Optimized amorphous Se (*a*-Se) generated from a monoclinic crystalline Se (*c*-Se) with computational quenching. The structure is dominated by disintegrated forms of Se rings. **(a2)** Radial Distribution Function (RDF) plot for monoclinic Se (*c*-Se) with nearest neighboring distance of ~2.4Å. More than one prominent peak is symbolic of crystallinity. **(b2)** RDF plot for amorphous Se (*a*-Se) obtained after quenching of monoclinic Se. Nearest neighboring distance is ~2.38Å and only one prominent peak is noted with low intensity. RDF plots for *c*-Se and *a*-Se conform and differentiate the structures of optimized Se allotropes. **(a3)** Representation of initial structure of *c*-Se/graphene interface prior to the interface study. **(b3)** Representation of initial structure of *a*-Se/graphene interface.

## 3. Results and Discussions

### 3.1 Interface Strength Analysis

To evaluate the strength of Se-Gr interfaces, we first computed *work of separation* ($W_{sep}$) for each interface system. By definition, it is the energy per unit area required to separate the two materials completely in the direction normal to the interface. To accomplish this, slab models for (*a*-/*c*-) Se/Gr were created with vacuum in z-dimension to permit atomic relaxation, as shown in fig.2-a,b. The standard description of $W_{sep}$ is as follows

$$W_{sep} = \sigma_1 + \sigma_2 - \gamma_{12} = \frac{E_1 + E_2 - E_{12}}{A} \quad (1)$$

Here, $\sigma_1$, $\sigma_2$ are surface energy of both the materials, $\gamma_{12}$ is the interface energy, $E_1$ and $E_2$ are total energy of slab 1 and slab 2, respectively. $E_{12}$ is the total energy of interface system in slab 3. $A$ is the area of contact at the interface. Besides Se/Gr, we also used similar slab models to calculate $W_{sep}$ in *c*-Se/Al and *a*-Si/Gr interface systems.

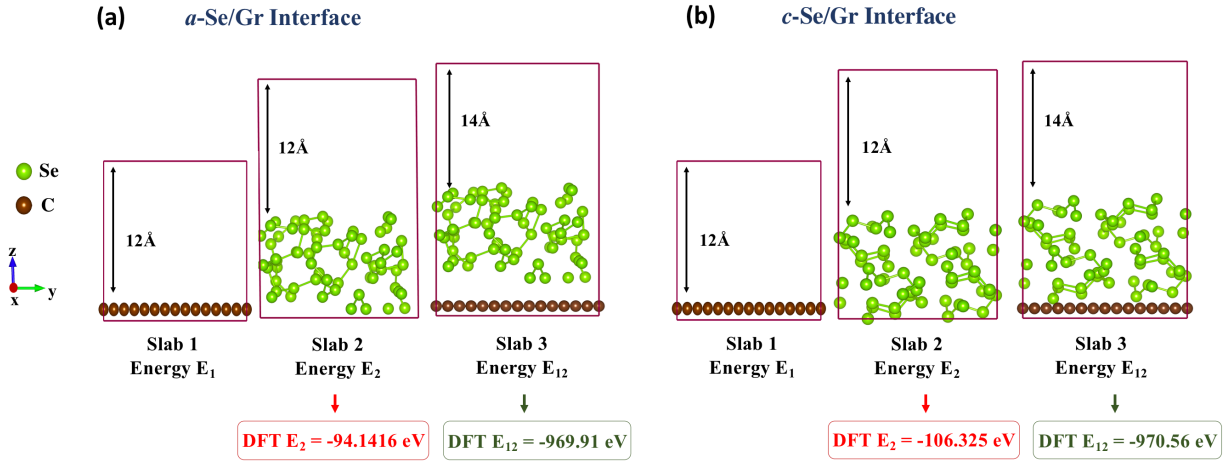

**Figure 2 Vacuum interface models with three slabs taken for the work of separation calculations. (a)** Three slabs taken for surface energy calculations of amorphous Se and Gr interface (*a*-Se/Gr). **(b)** Three slabs taken for surface energy calculations of monoclinic Se and Gr interface (*c*-Se/Gr). Vacuum of 12Å was added in z direction for slab 1 and 2, 14Å vacuum in z direction for slab 3 supercell containing the interface. **E$_2$** for *a*-Se is higher than for *c*-Se denoting the lower thermodynamical stability of amorphous Se phase. The **E$_{12}$** of both the interface systems (*c*-Se/Gr and *a*-Se/Ge) is almost same.

Results of W$_{sep}$ for different interfaces are summarized in fig. 3a and indicate that interface strength for Se-Gr systems (both *a*- and *c*-Se) is comparable to *a*-Si/Gr interface. Lower W$_{sep}$ has been previously shown[12] to influence electrode performance positively by mitigating stresses in Si electrode during lithiation/delithiation cycle. Computational analyses, backed by experimental validation, suggest that W$_{sep}$ value of ~ 0.41 J/m$^2$ (green interface in fig. 3c) for amorphous Si over Gr (*a*-Si/Gr) permits a 'slippery' vdW interface where Si is loosely physisorbed on Gr surface without any strong bonding. This allows these two materials to slip over one another in a frictionless manner without losing the mechanical contact. In contrast, the high W$_{sep}$ (~ 1.6 J/m$^2$) in amorphous Si over Ni (*a*-Si/Ni) is associated with 'non-slippery' high adhesion conditions dominated by repeated compression and tension in the interfacial region (red interface in fig.3b). In our study, W$_{sep}$ values for *a*-Se/Gr and *c*-Se/Gr are 0.34 J/m$^2$ and 0.43 J/m$^2$, respectively. The comparable interface strength of *c*-Se/Gr and *a*-Si/Gr propose long cycle life of Se-Gr electrodes. In addition, monoclinic *c*-Se and Gr interface system is devoid of any lattice mismatch associated lattice distortions. The 8-membered rings of Se are mostly conserved in the stable interface system with Gr. Upon optimization, there is only a slight vertical condensation (can be noted in fig. 4b) of Se crystal, resulting in minor distortions of dihedral angles and low interfacial gap (*d*), which works in favor of the interface in establishing a beneficial contact with Gr.

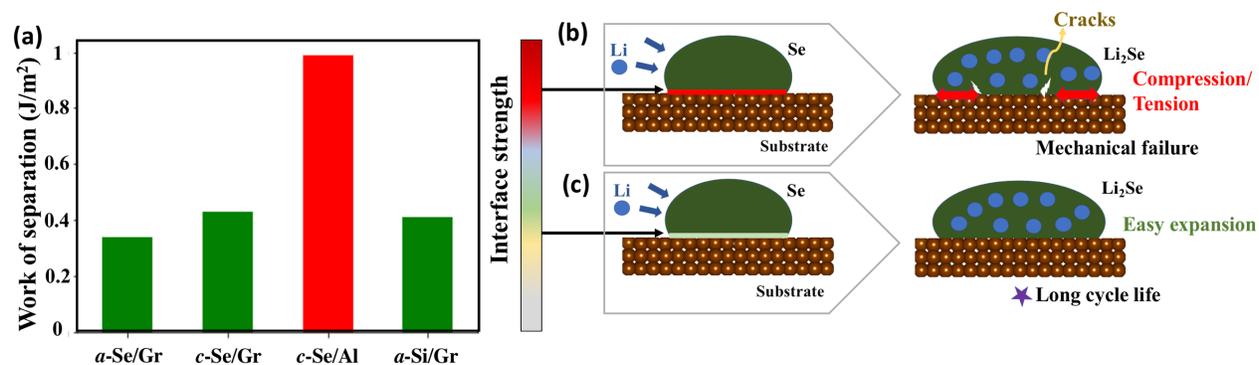

**Figure 3 Interface strength quantified by work of separation ($W_{sep}$) results. (a)** Interfacial work of separation for relaxed *a*-Se/Gr, *c*-Se/Gr, *c*-Se/Al and *a*-Si/Gr interfaces. **(b)** Schematic representation of an interface under 'high interface strength' condition denoted by red color, facing compressive stresses during Li incorporation in LIBs ultimately leading to crack propagation and mechanical failures. **(c)** Schematic representation of a contrasting 'low interface strength' condition as seen in the case of graphene interfaces and denoted by green/yellow here. Passive interface strength permits easy expansion and contraction to the active electrode material.

As alloying electrodes undergo continuous phase changes during battery cycle, Se will have an added advantage of similar interface strength during its phase transitions (*c*-Se ↔ *a*-Se) as compared to its complementary electrodes. $W_{sep}$ of *a*-Se/Gr (0.34 J/m$^2$) is less only by 20% of *c*-Se/Gr (0.43 J/m$^2$), primarily due to the structural similarities between the two phases. This was because the *a*-Se derived by the quenching process was similar to *c*-Se in terms of first neighboring Se-Se distances (~2.4Å in fig.1-a2,b2). The critical difference between the two allotropes of Se is that in *a*-Se, Se$_8$ rings break to form different sized polymeric chains, as shown in fig.1-b1. Our structures are in tune with previous studies on Se allotropes and confirms that *a*-Se is dominated by large chain molecules having each Se atom surrounded by two immediate neighbors, with interatomic distances similar to *c*-Se[44, 55]. In our calculations, the interface energy ($\gamma_{12} = E_{12}/A$) of both the interface systems (*c*-Se/Gr and *a*-Se/Ge) is almost same. However, the overall interface strength ($W_{sep}$) drops in *a*-Se/Gr system (yellow interface in fig.3) due to comparatively high system energy ($E_2$) and lower thermo-dynamical stability of *a*-Se phase as pointed out in fig 2. The disintegrated forms of Se rings dominate the *a*-Se/Ge structure.

In addition to the cycle life and phase transition, lower interface strength between Se and Gr can be beneficial in designing the electrode morphology. A latest study[43] shows high adhesion between active electrode material (AM) and binder causes disconnected lumps of AM-binder within the electrode. Passable interface strength between the active Se electrode and Gr flakes (as the binder) permits both the materials to be completely dispersed throughout the volume maintaining ionic and electronic conductive pathways. To present a contrast, interface strength in *c*-Se/Al system was examined by evaluating $W_{sep}$. The replacement of 2D Gr by a 3D Al affected the interface strength with a two-fold increase (0.99 J/m$^2$ red interface in fig.3). Al is also a conventional current collector used at cathode end in LIBs, and our results suggest that by reducing the surface contact between Se and Al, cycle life of Se electrode can be enhanced. This contrasting adhesion of Se with Gr and Al advocates the use of Se-Gr electrodes in battery applications. Next, we investigate compelling factors that contribute to the interfacial strength in Se and Si interface systems.

*3.2 Electron Exchange and Charge Separation analysis*

The low interface strength and slippery surface of Gr pose an essential question - how long does the Se-Gr interface stay intact? Se is previously reported to peel off from SiO$_2$/Si surfaces by slight mechanical exertion due to a lack of mechanical interlocking and chemical interaction[56]. This condition was improved by inserting an inconsistent intermediate layer of Indium(In) between Se-SiO$_2$ interface. A non-metal like Se could then be held in place by forming a surface alloy of In$_2$Se$_3$. In the case of 2D materials such as Gr, even with Se-Gr interface strength being similar to Si-Gr interface, Se-Gr might still lack stability due to polarity and absence of dangling bonds as prevalent in the case of Si[57, 58]. Applicatory longevity of Se interfaces needs to be further investigated by utilizing a comprehensive analysis of bonding. In this section, we discuss the persistence of Se-Gr interfaces as the function of electron distribution across the interface. We analyze electron exchange between the two surfaces, followed by the difference in surface potential and resultant charge separation between the two materials.

Electron redistribution is a prominent reason for interface strength and can throw light on the bonding phenomenon at the interface. The overall electron exchange between 3D Se bulk and 2D Gr is studied in optimized interface supercells via Bader charge analysis using scripts by Henkelman group[59]. Bader charge analysis quantifies atomic charges based on the charge density in a bader volume of each atom in the relaxed structure and calculates net charge transfer across the interface (**Δq**). In light of our used pseudopotential, all the C, Si, Se, and Al atoms in the system were taken to have 4, 4, 6, and 3 valence electrons, respectively. Charge distribution on Gr was computed by summing electronic charges on all the carbon atoms in the system (**$q_c$**). Then the total charge transfer across the interface was calculated by

$$\Delta q = q_c - 4 \times c \qquad (2)$$

where *c* is the number of carbon atoms in the system. The resultant values are presented in Table 1 for all considered Gr interface systems. The positive value of Δq indicates the number of electrons Gr gained when in contact with the bulk material, while a negative value represents the loss in electrons. The relation between the interface strength and the total charge transfer across the interface, $W_{sep} \propto |\Delta q|/d^2$, for Se interfaces compares well with some previous works on Pt-Gr and Si-C interfaces[51, 60]. In *a*-Se/Gr, *a*-Se gains net ~0.2556 electrons from Gr, leading to *p*-type doping in the latter. We observed that in *a*-Se structure, Se atoms, broken from the chains, adsorb on Gr surface by gaining more electrons (illustrated in fig.4a). This result is consistent with a previous work by Nakada et al.[61] which highlights while most atoms lose electrons to Gr surface, non-metals from Group 16 and 17 take up electrons from Gr. Therefore, Se atom gains about 0.01e$^{-1}$ when adsorbed on Gr surface. Our results verify that in an amorphous state, Se atoms adsorb on Gr surface with similar characteristics.

Direction of charge transfer is reversed in the crystalline interface system (fig. 4b), where *c*-Se loses electrons to Gr (~0.312e$^{-1}$). Atoms on Se$_8$ in the interfacial regions have less electrons than the atoms in Se$_8$ farther from Gr. The tendency of Gr to gain electrons from interfacing 3D bulk is steady in *a*-Si/Gr system where net 0.4226e$^{-1}$ is gained by Gr (summarized in Table 1). Additionally, a very distinctive interface is noted between *c*-Se and Al in fig. 4c, where Se$_8$ rings at the interface break into individual atoms to form strong covalent bonds with Al surface. There is a surface reaction between Al and Se surface atoms resulting in Δq = 4.5 e$^{-1}$ between Se and Al substrate. This reaction between Se and Al will result in the loss of active Se for reaction with oncoming adatom in batteries (e.g., Li in Lithium batteries).

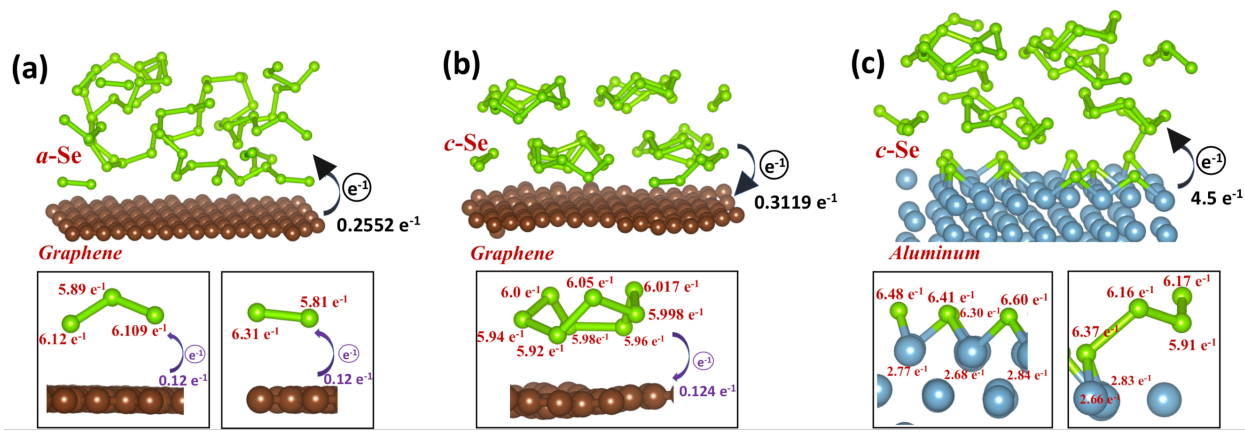

**Figure 4 Distribution of electrons on Se atoms present at the graphene and aluminum interfaces**. **(a)** Illustration of net charge transfer (Δq = 0.2552 e$^{-1}$) from Gr surface to *a*-Se at *a*-Se/Gr interface. Se atoms detached from Se chains and attached to fewer than 2-3 Se atoms, adsorb on Gr surface by gaining more electrons (~0.12 e$^{-1}$). **(b)** The net charge transfer (Δq = 0.3119 e$^{-1}$) at *c*-Se/Gr interface is directed towards Gr. Se atoms within Se$_8$ rings in the interfacial region have lower number of electrons than Se atoms farther from Gr surface. **(c)** Optimized view of *c*-Se/Al interface where Se$_8$ rings at the interface break into individual atoms to form covalent bonds with surface Al atoms. This surface reaction between Al and Se results in comparatively high net charge transfer (Δq = 4.5 e$^{-1}$) between Se and Al substrate. All the charges were obtained via Bader charge analysis.

Electron exchange (Δq) at Si-Gr interface is quantitatively more than Se-Gr interfaces (Table 1). This comparative ease of electron exchange at Si-Gr interface can be understood with the potential gradient and charge separation analysis presented in fig. 5. To bridge the electronic character across Se-Gr interface, we mapped the potential step (ΔV) between two materials and defined it as potential gradient (dϕ/dz) by dividing the difference in electrostatic potential at the interface with the interface gap (*d*). The computed electrostatic potential (V) on atoms was averaged in x-y plane for every unit z dimension (normal to Gr plane)[62]. Potential of Se and Gr at the interface were acquired by averaging $V_{Se/Gr}$ in the z dimension[63]. Potential gradient across the interface was determined by

$$\frac{d\phi}{dz} = \frac{V_{Se} - V_{Gr}}{d} \qquad (4)$$

where $V_{Se}$-$V_{Gr}$ is the difference of V between Se and Gr atoms at the interface. The interfacial gap (*d*) in the z dimension is denoted by the distance of lowest Se atom from the Gr surface (see fig.5-a1,b1,c1).

Lower potential gradient promises ease of interaction at the interface ($d\phi/dz$ ~0 for same materials), while large potential step is indicative of incohesive interface with less scope for electron exchange and bonding. The $d\phi/dz$ values for Se and Si interfaces with Gr are summarized in Table 1, along with their associated $W_{sep}$ and electron exchange results. Fig.5-a1,b1 demonstrates the potential step that developed across the Se-Gr interfaces and the resultant gradients. The red curve is the *averaged electrostatic potential* in the x-y plane, and purple is the averaged V across the z dimension. The difference in average potential of *a*-Se and Gr in the interface system is the highest (fig.5-a1). This results in a sizeable potential step and a steep value of $d\phi/dz$ (3.081 eV/Å). A similar trend is noted in the case of *c*-Se/Gr interface in fig.5-b1, where $d\phi/dz$ value is 3.03eV/Å and *d* = 2.86Å. This curtailed value of $d\phi/dz$ and *d* indicate *c*-Se/Gr interface system might be slightly superior to its amorphous counterpart in terms of bonding ability. In comparison to Se-Gr interfaces, *a*-Si/Gr has reduced potential step across the interface and resulting $d\phi/dz$ (2.18 eV/Å) is significantly lower (fig.5-c1). Hence, we observe higher electron exchange at *a*-Si/Gr interface than *c*-Se/Gr despite having comparative $W_{sep}$ values. The $d\phi/dz$ values indicate that Se is less likely to remain bonded with Gr as compared to the case of Si/Gr.

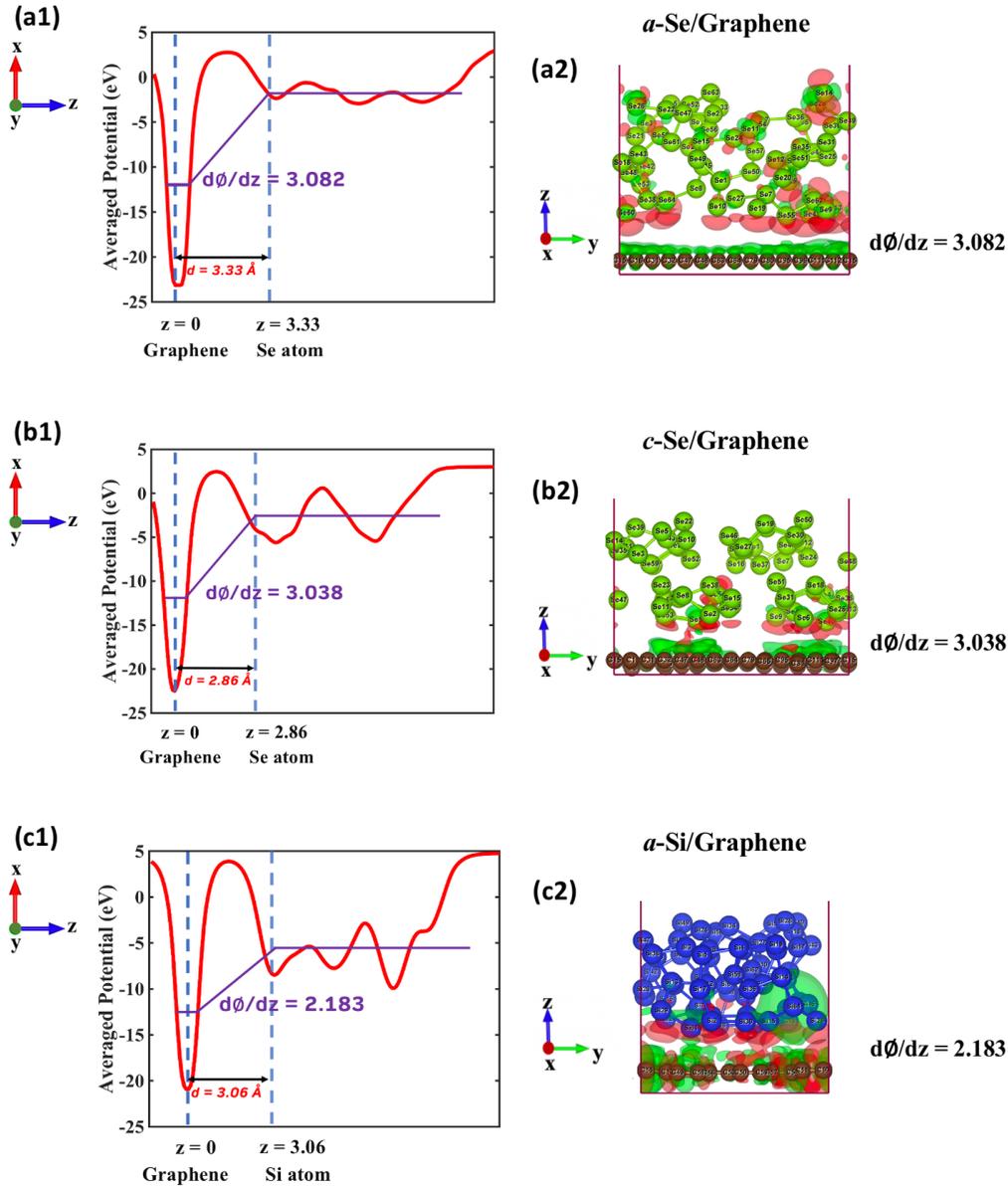

**Figure 5 Comparison of potential gradient and charge separation at graphene interfaces. (a1, b1, c1)** Averages potential curves at graphene interfaces with amorphous Se, crystalline Se and amorphous Si. The red curve is the averaged electrostatic potential in the x-y plane and purple is the averaged potential across the z dimension. There is a potential step across all the interfaces which results in potential gradients (d$\phi$/dz). *d* denotes the distance of the nearest Se/Si atom with respect to Gr sheet. **(a2, b2, c2)** Charge separation schemes for Gr interface with amorphous Se, crystalline Se and amorphous Si. Charge accumulation and depletion are shown in red and green, respectively. In comparison to Se-Gr interfaces, *a*-Si/Gr has reduced potential step (d$\phi$/dz = 2.18 eV/Å) and significant overlap of electron clouds across the interface representing ease of interfacial interaction. Large potential gradient (dϕ/dz) at Se/Gr interfaces is indicative of incohesive interface with less scope for electron exchange and bonding.

Charge density in the interfacial region was visualized by charge separation analysis. Charge separation scheme at the interface was extracted by subtracting charge density of individual materials from that of the entire system, and difference is plotted with an isosurface of 0.00024 e Å$^{-3}$. The resultant plots in fig.5-a2,b2,c2 provide the extent of interaction between the atomic systems and are consistent with our Δq and dϕ/dz results. Charge separation scheme for Se interfaces exhibits hardly any overlap of electron cloud between the two materials. Nevertheless, there is a presence of strong dipole at the interface due to accumulation of negative and positive charges, as indicated by red and green isosurfaces. Charge separation of *c*-Se/Gr (fig.5-b2) suggest the crystalline phase of Se is better than *a*-Se in forming a reliable interface with Gr as there is some overlap of positive and negative isosurfaces at the interface. Charge separation scheme of *a*-Si/Gr interface exhibits a better overlap of electron cloud between the two materials. These findings further imply that Se-Gr interfaces are not as amicable as Si-Gr, and Se alone can easily disintegrate from Gr surface upon external stimulation.

**Table 1** Summary of electron distribution across Gr interfaces with bulk amorphous Se, crystalline Se and amorphous Si along with its associated interface strength value (**W$_{sep}$**). The potential energy gradient across the interface is denoted by **dϕ/dz,** total charge transferred across the interface is given by **Δq,** where positive value denote charge acquired by graphene while negative value denote the charge given by graphene to the bulk, and *d* is the distance between Gr and lowest **Se/Si** atom.

| Systems | W$_{sep}$ (J/m$^2$) | dϕ/dz (eV/Å) | Δq (e$^{-1}$) | d (Å) |
|---|---|---|---|---|
| *a*-Se/Gr | 0.34 | 3.08 | -0.2552 | 3.33 |
| *c*-Se/Gr | 0.43 | 3.03 | 0.3119 | 2.87 |
| *a*-Si/Gr | 0.41 | 2.18 | 0.4266 | 3.06 |

*3.3 Electronic conductivity in Se-Gr interface systems*

In order to understand the differences in electronic conductivity of Se allotropes when interfaced with Gr, we have incorporated DOS analysis for *a*-Se, *c*-Se, and their respective Gr interface systems. DOS analysis gives an idea about the number of states electrons are allowed to occupy at a particular energy level in a system. In principle, the distribution of electronic states near the Fermi level (denoted by dark dashed line in fig. 6) is noted. Since our work focusses on how Gr interface changes for two different Se allotropes, we first analyzed DOS plots of *a*-Se and *c*-Se without a Gr substrate in fig.6-a,b. We find that the overall appearance of DOS changed between *a*-Se and *c*-Se. In particular, there is a significant reduction in band gap in total DOS plot of *a*-Se in comparison to *c*-Se. The band gap between valance band (VB) and conduction band (CB) in *c*-Se is 1.6 eV (fig.6b), which is reduced to 0.16 eV in *a*-Se (fig.6a). Moreover, energy states that are discrete for *c*-Se become more continuous in *a*-Se as existent peaks broaden and new energy levels are introduced due to the formation of disintegrated Se chains of different lengths. As such, the amorphous phase of Se looks more promising in terms of electronic conductivity.

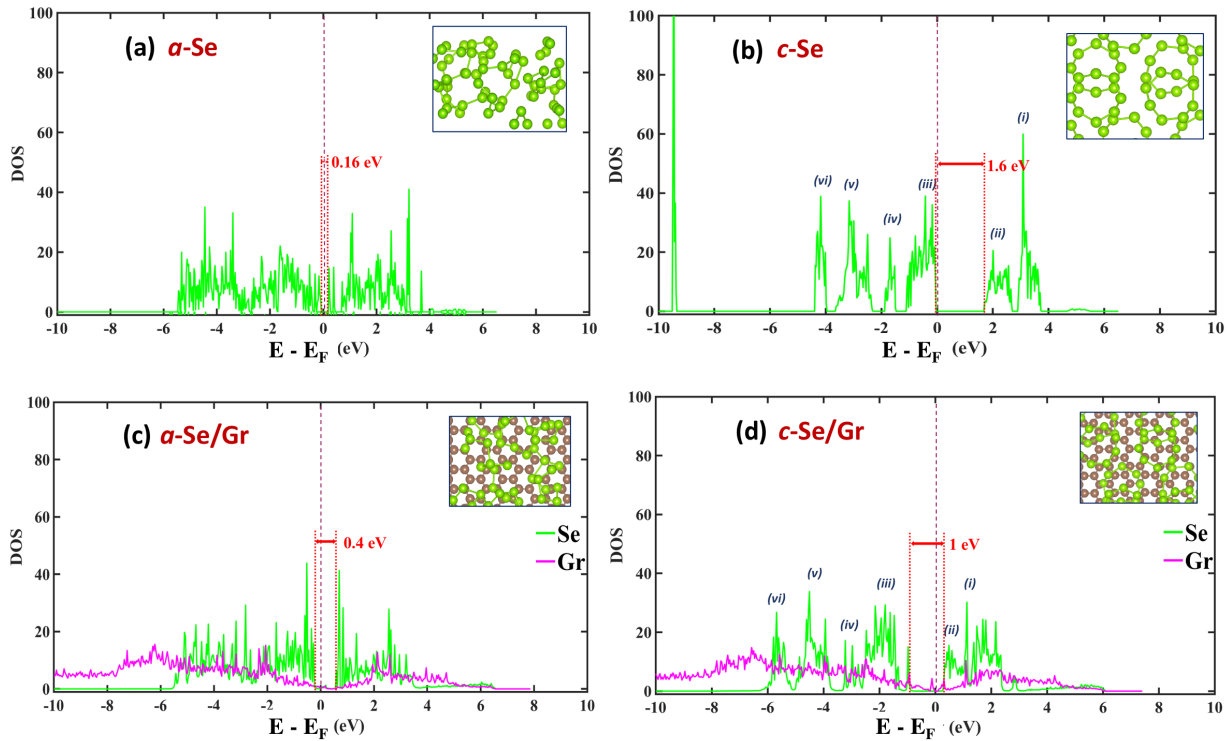

**Figure 6 Total Density of States (DOS) plots of Se allotropes and Se interfaced with graphene**. The green and pink plot indicates the total DOS of Se and Gr in the system **(a)** DOS of isolated amorphous selenium (*a*-Se). Energy states are continuous with very small band gap of 0.16 eV. **(b)** DOS of isolated monoclinic selenium (*c*-Se). Energy states are discrete, and peaks are labeled as *(i-vi)*. There is a difference of 1.6 eV between Valance band (VB) and Conduction band (CB). **(c)** Presents t-DOS of amorphous

selenium over graphene substrate (*a*-Se/Gr). Difference between VB and CB in *a*-Se increase to 0.4 eV. **(d)** Presents t-DOS of monoclinic selenium over graphene substrate (*c*-Se/Gr). All the labeled peaks shift towards lower energy as energy difference between VB and CB is reduced to 1 eV.

Next, we plotted total DOS for Se and Gr when interfaced in *a*-Se/Gr and *c*-Se/Gr systems (see fig.6-c,d) to interpret the influence of Gr substrate on electronic properties of Se and vice versa. In *c*-Se/Gr system, the presence of Gr with *c*-Se works towards slightly improving the conductivity in Se. First, the band gap in total DOS of Se (green), which was 1.6 eV before reduced to 1 eV. Distinct peaks of *c*-Se become broader in *c*-Se/Gr to show explicit continuity of energy states in VB and CB (fig.6d). Next, a redshift (towards lower energy) of all labeled peaks of *c*-Se is seen due to presence of Gr. This shift of peaks is on account of changes observed in dihedral angles of relaxed *c*-Se structure over Gr surface. In addition, some additional peaks are introduced due to overlap of selenium's 3d orbitals and carbon's p orbitals. Effects of Gr substrate on electronic properties of Se are slightly reversed in the case of *a*-Se/Gr. In contrast to its crystalline allotrope, presence of Gr brought about redistribution of states near the band gap ( compare fig.6 a and c). A recession of CB and VB is noted, which has introduced a band gap in the total DOS of *a*-Se in fig.6c in comparison to fig 6a. The energy between CB of *a*-Se, occupied by 3d electrons, and VB dominated by p orbital electrons, increases from 0.16 eV to 0.4 eV due to interference of Gr orbitals with Se. Overall, the continuity of electronic states in total DOS of *a*-Se/Gr indicates enhanced conductivity.

The presence of surface Se also impact the electronic properties of Gr as demonstrated in total DOS plots of pristine and Se-doped Gr in fig.7. Gr is a semi-metal known for its characteristic cone that conjoins VB and CB at the minimum conductivity point also known as Dirac point at 0 eV[64]. However, due to the presence of *c*-Se, the total DOS curve of Gr in *c*-Se/Gr deviates from its signature cone structure with incorporation of some additional states near 0 eV as it gains electrons. Most importantly, very small band gap of 0.10 eV is introduced near 0 eV in electronic states of Gr interfaced with monoclinic Se (see fig.7c). The surface charge states of Gr resultant of π electrons are known to be sensitive to surface charge distribution. We suggest that spread of $Se_8$ rings of monoclinic Se on the surface of Gr causes inconsistent charge densities on Gr surface (can also be seen in fig.5-b2). This redistribution of surface charges is the potential reason for opening of band gap in Gr. This band gap is desirable for device applications in order to control the behavior of charge carriers and hence can be effectively engineered based on the characteristics of the chemical dopant (active *c*-Se in this case). In contrast, graphene DOS maintains its distinctive conical structure in *a*-Se/Gr as shown in fig.7b. Through the loss of electrons to Se, the minimum conductivity point in Gr shifts towards positive gate voltage 0.4 eV from 0 eV. This shift is similar to the DOS of *p*-type doped Gr in previous studies[65]. Due to structural inconsistencies in the amorphous state of Se, the count of dopant acceptor Se atoms can vary which will influence the mobility of charge carriers in Gr. Besides *a*-Se having better electronic conductivity than its crystalline counterpart, *a*-Se/Gr can also have enhanced electron mobility. As the electron concentration increases on Gr due to oncoming donor metal (Li, Na, K, etc. in batteries), mobility of electron will tend to decrease. Free acceptor Se atoms at *a*-Se/Gr interface function to modulate electron concentrations. Overall, the total DOS plot of Gr in *c*-Se/Gr and *a*-Se/Gr, resonates well with our bader charge analysis (Δq) and charge separation results.

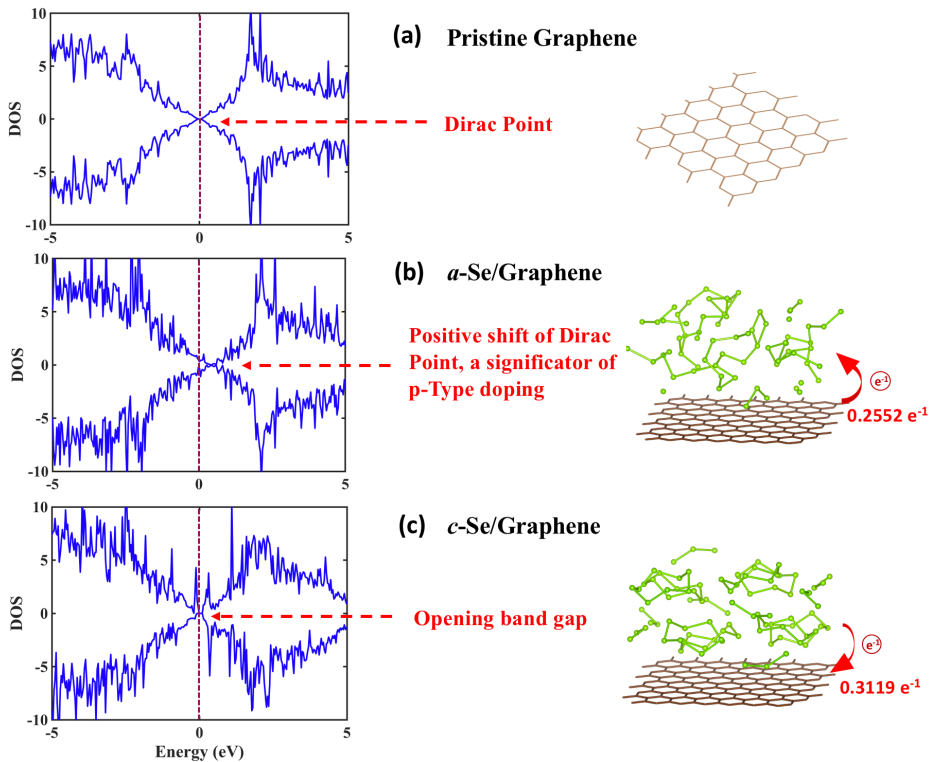

**Figure 7 Changes in total DOS of Graphene. (a)** DOS of pristine graphene having Dirac point at 0 eV where VB and CB meet. **(b)** Total DOS of graphene when interfaced with amorphous Se (*a*-Se). Due to surface charge doping of graphene by *a*-Se, Dirac point shifts towards 0.4 eV. **(c)** Total DOS of graphene when interfaced with monoclinic Se (*c*-Se) deviates from its signature cone structure with incorporation of some additional states near 0 eV as it gains electrons and introduction of a very small band gap of 0.10 eV near 0 eV.

## 4. Conclusions

In conclusion, we performed a comparative study of interfacial characteristics for Se-Gr interface and distinguished Si-Gr interface. By first principle calculations, we probed two Se-Gr interfaces with different Se allotropes for strength, long term stability and electronic conductivity. Our results show that

(i) Among two allotropes of Se, crystalline Se promises superior interface with Gr in terms of strength and persistence. Backed by amicable work of separation ($W_{sep}$ = 0.43 J/m$^2$) and fair interaction of charges at the interface ($\Delta q$ = 0.3119 e$^{-1}$), *c*-Se/Gr interface exhibits characteristics similar to *a*-Si/Gr ($W_{sep}$ = 0.41 J/m$^2$ and $\Delta q$ = 0.4266 e$^{-1}$).

(ii) Based on the quantitative evidence ($d\phi/dz$ and $\Delta q$) presented in our study, we doubt longevity of *c*-Se/Gr interface in comparison to *a*-Si/Gr. The potential difference between *c*-Se/Gr interface ($d\phi/dz$ = 3.03 eV/ Å) is larger than *a*-Si/Gr case ($d\phi/dz$ = 2.18 eV/ Å). In addition, the higher electron exchange is noted for *a*-Si/Gr interface ($\Delta q$ = 0.4266 e$^{-1}$) than *c*-Se/Gr ($\Delta q$ = 0.3119 e$^{-1}$) despite having comparative $W_{sep}$.

(iii) Even though, interface strength of *a*-Se/Gr ($W_{sep}$ = 0.34 J/m$^2$) differs only slightly from *c*-Se/Gr system ($W_{sep}$ = 0.43 J/m$^2$), there is significant anomaly in interfacial bonding. *c*-Se donates electrons to Gr. In *a*-Se, Se atoms are separated from polymeric chains and adsorb on Gr by acquiring electrons. This difference in interfacial electron exchange for Se allotropes influence the electronic character of both the Se-Gr systems as seen in our analysis of electronic conductivity using the DOS analysis.

(iv) The DOS plots of Gr interfaced with amorphous and crystalline Se denote *p*- and *n*- type doping of Gr, respectively. Therefore, as the mechanism of bonding is different for amorphous and crystalline Se with Gr, they activate Gr differently for wide range of catalytic and electrochemical applications.

Overall, our computational results provide deeper insight into the interface between Selenium and graphene.

## AUTHOR INFORMATION

### Corresponding Authors

*Email : dibakar.datta@njit.edu

*Email : david.mitlin@austin.utexas.edu

### Author Contributions
All authors have given approval to the final version of the manuscript

### Notes
The authors declare no competing financial interest.

## ACKNOWLEDGEMENT

D.M. (co-conception and guidance of research, preparation of manuscript) was supported by supported by the National Science Foundation, Civil, Mechanical and Manufacturing Innovation (CMMI), Award Number 1911905. V.S. (computation, manuscript preparation), D.D. (co-conception and guidance of research, preparation of manuscript) were supported by NSF (Award Number 1911900). D.D. acknowledges Extreme Science and Engineering Discovery Environment (XSEDE) for the computational facilities (Award Number – DMR180013). V.S. acknowledges Dr. Kamalika Ghatak for the fruitful discussion during the project.

## DATA AVAILABILITY

The data reported in this paper is available from the corresponding author upon reasonable request.

## CODE AVAILABILITY

The pre- and post-processing codes used in this paper are available from the corresponding author upon reasonable request. Restrictions apply to the availability of the simulation codes, which were used under license for this study.

## REFERENCES

1. Datta, D.; Li, J.; Koratkar, N.; Shenoy, V. B., Enhanced lithiation in defective graphene. *Carbon* **2014,** *80*, 305-310.
2. Buldum, A.; Tetiker, G., First-principles study of graphene-lithium structures for battery applications. *Journal of Applied Physics* **2013,** *113* (15), 154312.
3. Chou, S.-L.; Wang, J.-Z.; Choucair, M.; Liu, H.-K.; Stride, J. A.; Dou, S.-X., Enhanced reversible lithium storage in a nanosize silicon/graphene composite. *Electrochemistry Communications* **2010,** *12* (2), 303-306.
4. Luo, J.; Zhao, X.; Wu, J.; Jang, H. D.; Kung, H. H.; Huang, J., Crumpled graphene-encapsulated Si nanoparticles for lithium ion battery anodes. *The journal of physical chemistry letters* **2012,** *3* (13), 1824-1829.
5. Wen, Y.; Zhu, Y.; Langrock, A.; Manivannan, A.; Ehrman, S. H.; Wang, C., Graphene-bonded and-encapsulated Si nanoparticles for lithium ion battery anodes. *Small* **2013,** *9* (16), 2810-2816.
6. Ji, J.; Ji, H.; Zhang, L. L.; Zhao, X.; Bai, X.; Fan, X.; Zhang, F.; Ruoff, R. S., Graphene-encapsulated Si on ultrathin-graphite foam as anode for high capacity lithium-ion batteries. *Advanced Materials* **2013,** *25* (33), 4673-4677.
7. Ko, M.; Chae, S.; Jeong, S.; Oh, P.; Cho, J., Elastic a-silicon nanoparticle backboned graphene hybrid as a self-compacting anode for high-rate lithium ion batteries. *ACS nano* **2014,** *8* (8), 8591-8599.
8. Zhao, M.-Q.; Zhang, Q.; Huang, J.-Q.; Tian, G.-L.; Nie, J.-Q.; Peng, H.-J.; Wei, F., Unstacked double-layer templated graphene for high-rate lithium–sulphur batteries. *Nature communications* **2014,** *5* (1), 1-8.
9. Zhou, G.; Pei, S.; Li, L.; Wang, D. W.; Wang, S.; Huang, K.; Yin, L. C.; Li, F.; Cheng, H. M., A graphene–pure-sulfur sandwich structure for ultrafast, long-life lithium–sulfur batteries. *Advanced materials* **2014,** *26* (4), 625-631.
10. Datta, M. K.; Epur, R.; Saha, P.; Kadakia, K.; Park, S. K.; Kumta, P. N., Tin and graphite based nanocomposites: Potential anode for sodium ion batteries. *Journal of Power Sources* **2013,** *225*, 316-322.
11. Yang, Y.; Zheng, G.; Cui, Y., Nanostructured sulfur cathodes. *Chemical Society Reviews* **2013,** *42* (7), 3018-3032.
12. Basu, S.; Suresh, S.; Ghatak, K.; Bartolucci, S. F.; Gupta, T.; Hundekar, P.; Kumar, R.; Lu, T.-M.; Datta, D.; Shi, Y., Utilizing van der Waals slippery interfaces to enhance the electrochemical stability of Silicon film anodes in Lithium-ion batteries. *ACS applied materials & interfaces* **2018**.

13. Rana, K.; Singh, J.; Lee, J.-T.; Park, J. H.; Ahn, J.-H., Highly conductive freestanding graphene films as anode current collectors for flexible lithium-ion batteries. *ACS applied materials & interfaces* **2014,** *6* (14), 11158-11166.
14. Park, M.-S.; Rajendran, S.; Kang, Y.-M.; Han, K.-S.; Han, Y.-S.; Lee, J.-Y., Si–Ni alloy–graphite composite synthesized by arc-melting and high-energy mechanical milling for use as an anode in lithium-ion batteries. *Journal of power sources* **2006,** *158* (1), 650-653.
15. Schmuelling, G.; Winter, M.; Placke, T., Investigating the Mg–Si binary system via combinatorial sputter deposition as high energy density anodes for lithium-ion batteries. *ACS applied materials & interfaces* **2015,** *7* (36), 20124-20133.
16. Datta, M. K.; Kumta, P. N., Silicon and carbon based composite anodes for lithium ion batteries. *Journal of Power Sources* **2006,** *158* (1), 557-563.
17. Shu, J.; Li, H.; Yang, R.; Shi, Y.; Huang, X., Cage-like carbon nanotubes/Si composite as anode material for lithium ion batteries. *Electrochemistry Communications* **2006,** *8* (1), 51-54.
18. Lv, D.; Zheng, J.; Li, Q.; Xie, X.; Ferrara, S.; Nie, Z.; Mehdi, L. B.; Browning, N. D.; Zhang, J. G.; Graff, G. L., High energy density lithium–sulfur batteries: challenges of thick sulfur cathodes. *Advanced Energy Materials* **2015,** *5* (16), 1402290.
19. Mikhaylik, Y. V.; Akridge, J. R., c. *Journal of the Electrochemical Society* **2004,** *151* (11), A1969.
20. Pang, Q.; Liang, X.; Kwok, C. Y.; Nazar, L. F., Advances in lithium–sulfur batteries based on multifunctional cathodes and electrolytes. *Nature Energy* **2016,** *1* (9), 1-11.
21. Xu, G.; Ding, B.; Pan, J.; Nie, P.; Shen, L.; Zhang, X., High performance lithium–sulfur batteries: advances and challenges. *Journal of Materials Chemistry A* **2014,** *2* (32), 12662-12676.
22. Su, D.; Zhou, D.; Wang, C.; Wang, G., Lithium-Sulfur Batteries: Toward High Performance Lithium–Sulfur Batteries Based on Li2S Cathodes and Beyond: Status, Challenges, and Perspectives (Adv. Funct. Mater. 38/2018). *Advanced Functional Materials* **2018,** *28* (38), 1870273.
23. Zhang, J.; Yin, Y. X.; You, Y.; Yan, Y.; Guo, Y. G., A High-Capacity Tellurium@ Carbon Anode Material for Lithium-Ion Batteries. *Energy Technology* **2014,** *2* (9-10), 757-762.
24. Liu, Y.; Wang, J.; Xu, Y.; Zhu, Y.; Bigio, D.; Wang, C., Lithium–tellurium batteries based on tellurium/porous carbon composite. *Journal of Materials Chemistry A* **2014,** *2* (31), 12201-12207.
25. Seo, J.-U.; Seong, G.-K.; Park, C.-M., Te/C nanocomposites for Li-Te secondary batteries. *Scientific reports* **2015,** *5* (1), 1-7.
26. Ding, N.; Chen, S. F.; Geng, D. S.; Chien, S. W.; An, T.; Hor, T. A.; Liu, Z. L.; Yu, S. H.; Zong, Y., Tellurium@ Ordered Macroporous Carbon Composite and Free-Standing Tellurium Nanowire Mat as Cathode Materials for Rechargeable Lithium–Tellurium Batteries. *Advanced Energy Materials* **2015,** *5* (8), 1401999.
27. Yang, C. P.; Xin, S.; Yin, Y. X.; Ye, H.; Zhang, J.; Guo, Y. G., An advanced selenium–carbon cathode for rechargeable lithium–selenium batteries. *Angewandte Chemie International Edition* **2013,** *52* (32), 8363-8367.
28. Abouimrane, A.; Dambournet, D.; Chapman, K. W.; Chupas, P. J.; Weng, W.; Amine, K., A new class of lithium and sodium rechargeable batteries based on selenium and selenium–sulfur as a positive electrode. *Journal of the American chemical society* **2012,** *134* (10), 4505-4508.
29. Liu, L.; Hou, Y.; Wu, X.; Xiao, S.; Chang, Z.; Yang, Y.; Wu, Y., Nanoporous selenium as a cathode material for rechargeable lithium–selenium batteries. *Chemical communications* **2013,** *49* (98), 11515-11517.
30. Li, Q.; Liu, H.; Yao, Z.; Cheng, J.; Li, T.; Li, Y.; Wolverton, C.; Wu, J.; Dravid, V. P., Electrochemistry of selenium with sodium and lithium: kinetics and reaction mechanism. *ACS nano* **2016,** *10* (9), 8788-8795.
31. Osada, N.; Bucur, C. B.; Aso, H.; Muldoon, J., The design of nanostructured sulfur cathodes using layer by layer assembly. *Energy & Environmental Science* **2016,** *9* (5), 1668-1673.
32. Yang, X.; Zhang, L.; Zhang, F.; Huang, Y.; Chen, Y., Sulfur-infiltrated graphene-based layered porous carbon cathodes for high-performance lithium–sulfur batteries. *Acs Nano* **2014,** *8* (5), 5208-5215.
33. Zhang, Z.; Yang, X.; Guo, Z.; Qu, Y.; Li, J.; Lai, Y., Selenium/carbon-rich core–shell composites as cathode materials for rechargeable lithium–selenium batteries. *Journal of Power Sources* **2015,** *279*, 88-93.
34. Xue, P.; Zhai, Y.; Wang, N.; Zhang, Y.; Lu, Z.; Liu, Y.; Bai, Z.; Han, B.; Zou, G.; Dou, S., Selenium@ Hollow mesoporous carbon composites for high-rate and long-cycling lithium/sodium-ion batteries. *Chemical Engineering Journal* **2019**, 123676.
35. Zhou, J.; Chen, M.; Wang, T.; Li, S.; Zhang, Q.; Zhang, M.; Xu, H.; Liu, J.; Liang, J.; Zhu, J., Covalent Selenium Embedded in Hierarchical Carbon Nanofibers for Ultra-High Areal Capacity Li-Se Batteries. *Iscience* **2020,** *23* (3), 100919.
36. Xi, K.; Kidambi, P. R.; Chen, R.; Gao, C.; Peng, X.; Ducati, C.; Hofmann, S.; Kumar, R. V., Binder free three-dimensional sulphur/few-layer graphene foam cathode with enhanced high-rate capability for rechargeable lithium sulphur batteries. *Nanoscale* **2014,** *6* (11), 5746-5753.
37. Zhang, J.; Xu, Y.; Fan, L.; Zhu, Y.; Liang, J.; Qian, Y., Graphene–encapsulated selenium/polyaniline core–shell nanowires with enhanced electrochemical performance for Li–Se batteries. *Nano Energy* **2015,** *13*, 592-600.
38. Kundu, D.; Krumeich, F.; Nesper, R., Investigation of nano-fibrous selenium and its polypyrrole and graphene composite as cathode material for rechargeable Li-batteries. *Journal of power sources* **2013,** *236*, 112-117.
39. Han, K.; Liu, Z.; Shen, J.; Lin, Y.; Dai, F.; Ye, H., A Free-Standing and Ultralong-Life Lithium-Selenium Battery Cathode Enabled by 3D Mesoporous Carbon/Graphene Hierarchical Architecture. *Advanced Functional Materials* **2015,** *25* (3), 455-463.
40. Fan, X.-Y.; Ke, F.-S.; Wei, G.-Z.; Huang, L.; Sun, S.-G., Sn–Co alloy anode using porous Cu as current collector for lithium ion battery. *Journal of Alloys and Compounds* **2009,** *476* (1-2), 70-73.
41. Kang, S.; Xie, H.; Zhai, W.; Ma, Z.; Wang, R.; Zhang, W., Enhancing performance of a lithium ion battery by optimizing the surface properties of the current collector. *Int. J. Electrochem. Sci* **2015,** *10*, 2324-2335.
42. Jeon, H.; Cho, I.; Jo, H.; Kim, K.; Ryou, M.-H.; Lee, Y. M., Highly rough copper current collector: improving adhesion property between a silicon electrode and current collector for flexible lithium-ion batteries. *Rsc Advances* **2017,** *7* (57), 35681-35686.

43. Srivastava, I.; Bolintineanu, D. S.; Lechman, J. B.; Roberts, S. A., Controlling Binder Adhesion to Impact Electrode Mesostructure and Transport. *ECSarXiv. doi* **2019,** *10*.
44. Goldan, A.; Li, C.; Pennycook, S.; Schneider, J.; Blom, A.; Zhao, W., Molecular structure of vapor-deposited amorphous selenium. *Journal of Applied Physics* **2016,** *120* (13), 135101.
45. Johari, P.; Qi, Y.; Shenoy, V. B., The mixing mechanism during lithiation of Si negative electrode in Li-ion batteries: an ab initio molecular dynamics study. *Nano letters* **2011,** *11* (12), 5494-5500.
46. Sharma, V.; Ghatak, K.; Datta, D., Amorphous germanium as a promising anode material for sodium ion batteries: a first principle study. *Journal of Materials Science* **2018,** *53* (20), 14423-14434.
47. Kresse, G.; Furthmüller, J., Efficient iterative schemes for ab initio total-energy calculations using a plane-wave basis set. *Physical review B* **1996,** *54* (16), 11169.
48. Kresse, G.; Joubert, D., From ultrasoft pseudopotentials to the projector augmented-wave method. *Physical Review B* **1999,** *59* (3), 1758.
49. Blöchl, P. E., Projector augmented-wave method. *Physical review B* **1994,** *50* (24), 17953.
50. Perdew, J. P.; Burke, K.; Ernzerhof, M., Generalized gradient approximation made simple. *Physical review letters* **1996,** *77* (18), 3865.
51. Stournara, M. E.; Qi, Y.; Shenoy, V. B., From ab initio calculations to multiscale design of Si/C core–shell particles for Li-ion anodes. *Nano letters* **2014,** *14* (4), 2140-2149.
52. Dion, M.; Rydberg, H.; Schröder, E.; Langreth, D. C.; Lundqvist, B. I., Van der Waals density functional for general geometries. *Physical review letters* **2004,** *92* (24), 246401.
53. Klimeš, J.; Bowler, D. R.; Michaelides, A., Chemical accuracy for the van der Waals density functional. *Journal of Physics: Condensed Matter* **2009,** *22* (2), 022201.
54. Klimeš, J.; Bowler, D. R.; Michaelides, A., Van der Waals density functionals applied to solids. *Physical Review B* **2011,** *83* (19), 195131.
55. Escoffery, C.; Halperin, S., Studies on the Polymorphic Transformation of Selenium. *Transactions of The Electrochemical Society* **1946,** *90* (1), 163.
56. Chen, Y.-Z.; You, Y.-T.; Chen, P.-J.; Li, D.; Su, T.-Y.; Lee, L.; Shih, Y.-C.; Chen, C.-W.; Chang, C.-C.; Wang, Y.-C., Environmentally and mechanically stable selenium 1D/2D hybrid structures for broad-range photoresponse from ultraviolet to infrared wavelengths. *ACS applied materials & interfaces* **2018,** *10* (41), 35477-35486.
57. Tao, M.; Udeshi, D.; Basit, N.; Maldonado, E.; Kirk, W. P., Removal of dangling bonds and surface states on silicon (001) with a monolayer of selenium. *Applied physics letters* **2003,** *82* (10), 1559-1561.
58. Lannoo, M., The role of dangling bonds in the properties of surfaces and interfaces of semiconductors. *Revue de physique appliquée* **1990,** *25* (9), 887-894.
59. Sanville, E.; Kenny, S. D.; Smith, R.; Henkelman, G., Improved grid-based algorithm for Bader charge allocation. *Journal of computational chemistry* **2007,** *28* (5), 899-908.
60. Sen, F.; Qi, Y.; Alpas, A., Anchoring platinum on graphene using metallic adatoms: a first principles investigation. *Journal of Physics: Condensed Matter* **2012,** *24* (22), 225003.
61. Nakada, K.; Ishii, A., Migration of adatom adsorption on graphene using DFT calculation. *Solid State Communications* **2011,** *151* (1), 13-16.
62. Peressi, M.; Binggeli, N.; Baldereschi, A., Band engineering at interfaces: theory and numerical experiments. *Journal of Physics D: Applied Physics* **1998,** *31* (11), 1273.

63.	Li, D.; Guo, L.; Li, L.; Lu, H., Electron work function–a probe for interfacial diagnosis. *Scientific reports* **2017,** *7* (1), 1-8.
64.	Novoselov, K.; Morozov, S.; Mohinddin, T.; Ponomarenko, L.; Elias, D.; Yang, R.; Barbolina, I.; Blake, P.; Booth, T.; Jiang, D., Electronic properties of graphene. *physica status solidi (b)* **2007,** *244* (11), 4106-4111.
65.	Tang, Y.-B.; Yin, L.-C.; Yang, Y.; Bo, X.-H.; Cao, Y.-L.; Wang, H.-E.; Zhang, W.-J.; Bello, I.; Lee, S.-T.; Cheng, H.-M., Tunable band gaps and p-type transport properties of boron-doped graphenes by controllable ion doping using reactive microwave plasma. *Acs Nano* **2012,** *6* (3), 1970-1978.